\documentstyle[twocolumn,aps]{revtex}
\begin{document}
\draft
\title{Flux penetration and expulsion in thin superconducting disks} 
\author
{V.A. Schweigert \cite {*:gnu} and F.M. Peeters \cite {f:gnu}}
\address{\it  Departement Natuurkunde, Universiteit Antwerpen (UIA),\\
Universiteitsplein 1, B-2610 Antwerpen, Belgium}
\date{\today}
\maketitle
\begin{abstract}
Using an expansion of the order parameter over the eigenfunctions of
the linearized first Ginzburg-Landau (GL) equation,
we obtain numerically the saddle points of the free energy separating the
stable states with different number of vortices.  
In contrast to known surface and geometrical barrier models, we find
that in a wide range of magnetic fields below the penetration field,
the saddle point state for flux penetration into a disk does not correspond
to a vortex located nearby the sample boundary, but to a
region of suppressed superconductivity at the disk edge with no winding of the
current, and which is {\it a nucleus} for the following vortex creation.
The height of this {\it nucleation barrier}, which determines the
time of flux penetration, is calculated for different disk radii
and magnetic fields.
\end{abstract}

The study of
magnetic flux penetration and expulsion in type-II superconductors
has traditionally attracted much attention in  view of important
technological and fundamental questions concerning hysteretic
behavior and phase transitions in bounded samples.
The vortex
creation problem is also related to phase transitions in superfluids.
It is well known \cite{gennes} that for type-II
superconductors ($\lambda/\xi>1/\sqrt{2}$, $\lambda$, $\xi$ -
are the penetration and coherence lengths, respectively),
  the Meissner state 
becomes energetically unfavorable
with increasing magnetic field at $H=H_{c1}$
 in comparison to the Abrikosov vortex lattice.
In a finite system these two states, which correspond to 
minima of the superconductor free energy, are separated by a barrier.
Therefore, a first-order transition between the Meissner and Abrikosov
states takes some time which 
decreases with temperature and 
approximately follows an Arrhenius law $\tau\propto exp(U/kT)$, where
$U$, $T$ are the barrier height and sample
temperature, respectively. For $T=0$, 
the Meissner state survives up to the penetration field $H_p$
and transits suddenly to the Abrikosov state due to dynamic
instability of the order parameter \cite{frahm}.
 With decreasing the magnetic field at zero temperature,
the vortex state remains stable up to the expulsion field $H_e<H_{c1}$
and then goes to the Meissner state due to vortex expulsion.

The origin of barriers for flux penetration and expulsion has been
discussed during the last thirty years. According to the
Bean-Livingston (BL)
model\cite{bean}, the {\it surface barrier}
appears due to a competition between the vortex
attraction to the sample walls by its mirror image
and its repulsion by screening currents.
This model was further developed for: i) cylindrical samples, where the
vortex shape was assumed not  to be an infinite line
but a semicircle \cite{semi},
ii) thin disks \cite{fetter} and iii) strips \cite{kogan}, where shielding,
due to finite size effects, does
not decay exponentially.
For samples with a non-elliptical cross section, the {\it geometrical barrier}
arises because of Meissner screening currents flowing on the top and
bottom surfaces of a flat strip \cite{geomtr}. In addition, 
{\it vortex pinning} by defects can play an important role in the delay
of vortex expulsion or promotion of vortex penetration. It should be stressed,
that the above mentioned  barrier models,
which are based on the London theory, do not account for the process of
vortex formation and describe only the vortex motion far
from the sample boundary.  The  Ginzburg-Landau (GL)
theory has been previously applied for the study
of barriers only for the 1D cases of narrow wires and rings\cite{langer}.
The approaches based on solving time-dependent GL equations \cite{frahm}
allow to treat flux penetration (expulsion) only for magnetic fields
higher (lower) than the penetration (expulsion) field.
In this Letter, starting from the non-linear GL theory
we present an approach  for finding the saddle point
states in thin disks
and calculate numerically the heights of the free energy barriers separating
the stable states with different number of vortices.

 We consider a superconducting defect free disk with radius $R$ and thickness
$d$ immersed in an insulating media
in the presence of a perpendicular uniform magnetic field $H$. 
For thin disks $Rd\ll \lambda^2$ we can neglect the distortion of the
magnetic field, which are induced by screening and vortex currents,
and write the GL functional as
\begin{equation}
\label{eq0}
G=G_n+\int d\vec r
\left(\alpha |\Psi|^2+\frac{\beta}{2}|\Psi|^4
+\Psi^*\hat L\Psi\right),
\end{equation}
where $G$, $G_n$ are the free energies of the superconducting and
normal states; $\Psi$ is the complex order parameter;
$\hat L=(-i\hbar\vec \nabla-e^{*}\vec A/c)^2/2m^*$ is the 
kinetic energy operator for Cooper pairs of charge
$e^{*}=2e$ and mass $m^{*}=2m$; $\vec A=\vec e_{\phi}H\rho/2$
is the vector potential of the uniform magnetic field written in
cylindrical coordinates $\phi,\rho$; 
 $\alpha$, $\beta$ are the GL coefficients
depending on the sample temperature.
Expanding the order parameter $\Psi=\sum_i^{N} C_i\psi_i$ in the
orthonormal eigenfunctions  of the
kinetic energy operator $\hat L\psi_i=\epsilon_i\psi_i$
 \cite{palacios,schweigert1,schweigert2}
we go from the functional form (\ref{eq0}) to the free energy
written in terms of complex variables
\begin{equation}
\label{gl2}
G-G_n=(\alpha+\epsilon_i)C_iC_i^*+\frac{\beta}{2}A^{ij}_{kl}C_i^*C_j^*C_kC_l,
\end{equation}
where the matrix elements $A^{ij}_{kl}=\int d\vec r\psi_i^*\psi_j^*\psi_k\psi_l$ are
calculated numerically.
Note, that the sample geometry enters in the
calculations only through the eigen energies $\epsilon_i$
and eigenfunctons $\psi_i$,
which are well known for the disk case \cite{schweigert1,moschchalkov}.
In thin ($d\ll \xi$) disks
these eigenfunctions have the form $\psi_{i=(l,n)}=exp(il\phi)f_n(\rho)$, where
$l$ is the angular momentum and the index $n$ counts
different states with the same $l$.
 In contrast to the approaches
 \cite{palacios,schweigert1,schweigert2}, we do not restrict ourselves
to the lowest Landau level approximation (i.e. $n=1$) and expand the order
parameter over all eigenfunctions with $\epsilon_i<\epsilon _{*}$,
where the cutting parameter $\epsilon_{*}$ is choosen such that
increasing it
does not influence our results. The typical number of complex
components used are  in the range $N=30\div50$. Thus the
superconducting state is mapped into
a 2D cluster of $N$ classical particles
$\left (x,y)\leftrightarrow (Re(C), Im(C)\right)$, which is governed
by the Hamiltonian (\ref{gl2}).

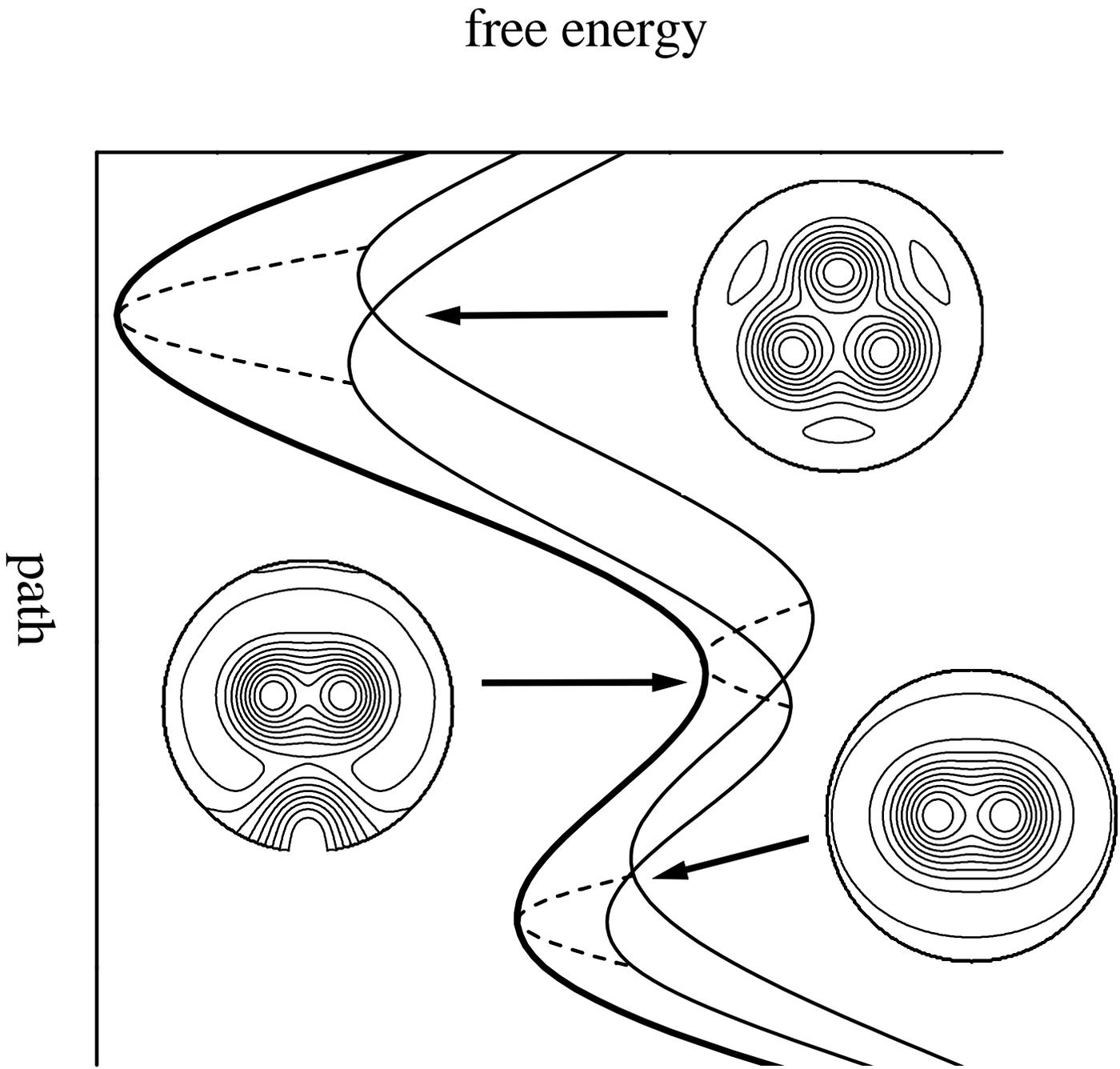
\begin{figure}
\caption{
Schematical view of the free energy in functional space depicting two
minima with $L=2$ and $L=3$ vortices and the saddle point connecting them.
The Cooper pair density of these three states are shown in the insets.
}
\label{fig1}
\end{figure}

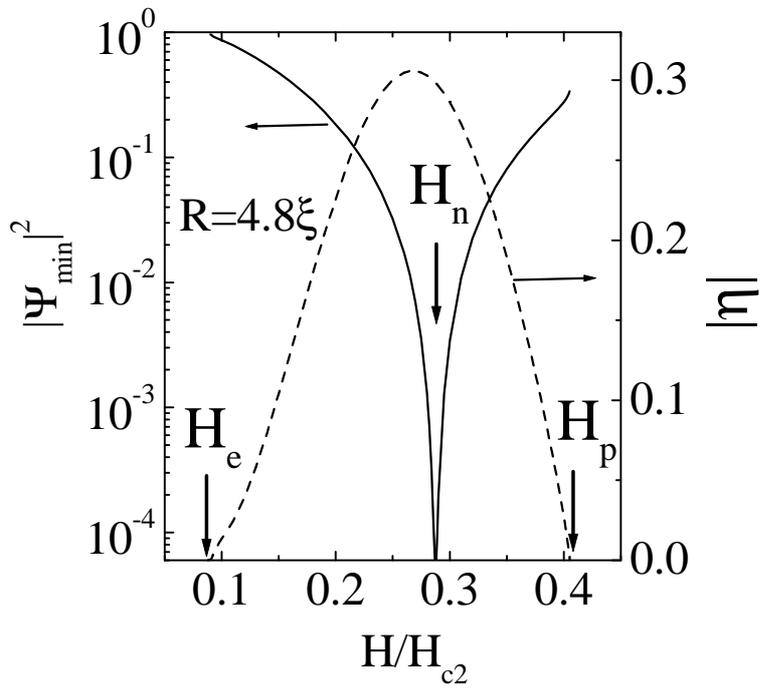
\begin{figure}
\caption{
The minimal density of superconducting electrons $|\Psi|^2$ at a
point of the
disk boundary (solid curve), measured in $-\alpha/\beta$, and the curvature
$\eta$ (dashed curve) of the free energy curve for the transition between the
Meissner and vortex states.
}
\label{fig2}
\end{figure}

To find a saddle point, which presents  an extremum of
the free energy, we use a technique similar to
the eigenvector following method \cite{eigen}.
We start with some set of  coefficients $C$. In the vicinity of this
point the free energy $\delta G=G(C^n)-G(C)$ can be represented as a quadratic
form for small deviations $\delta=C^n-C$:
\begin{equation}
\label{gl3}
\delta G=F_m\delta_m^*+B_{mn}\delta_n\delta_m^*
+D_{mn}\delta_n^*\delta_m^*+c.c.,
\end{equation}
where
$F_m=(\alpha+\epsilon_i)C_m+\beta A^{mj}_{kl}C_jC_k^*C_l$,
$B_{mn}=(\alpha+\epsilon_m)I_{mn}+2\beta A^{mn}_{kl}C_kC_l^*$,
$D_{mn}=\beta A^{mn}_{kl}C_kC_l$, and $I_{mn}$ is the unit matrix.
The quadratic form (\ref{gl3}), which is Hermitian, can
be rewritten in normal coordinates $\delta_m=x_kQ_m^k$ as
$\delta G=2(\gamma_kx_k+\eta_k x_k^2)$, where $\gamma_k=Q_m^kF_m$,
the eigenvalues
$\eta_k$ and
eigenvectors $Q^k$ are found by solving numerically the following equation
$$
\left |
\begin{array}{ll}
B+Re(D)&Im(D) \\
Im(D)&B-Re(D) 
\end{array}
\right |\cdot
\left |
\begin{array}{c}
Re(Q^k)\\
Im(Q^k) 
\end{array}
\right |=\eta_k
\left |
\begin{array}{c}
Re(Q^k)\\
Im(Q^k) 
\end{array}
\right |
 $$
Moving in the direction with negative free energy gradient $-\gamma_k$
we will approach a
minimum of the free energy corresponding to the ground or a metastable state.
In order to find a saddle point we
move to a minimum  of the free energy in all directions 
$x_{k\neq l}=-\gamma_k/(\epsilon+\eta_k)$ except one, which has
the lowest eigenvalue, and for which we go to a maximum
$x_{l}= \gamma_l/(-\epsilon+\eta_l)$, and find 
$C_m^n=C_m+x_kQ_m^k$ for all $k$.
The iteration parameter
$\epsilon>0$ controls the convergency, which is always reached
for any initial state close enough to a saddle point.
Starting from
different initial states, for which the coefficients $C$ are choosen
randomly, we find the saddle points for different magnetic fields
(Fig.~\ref{fig1}). Due to fluctuations (i.e. thermal, ...) the system will
be able to reach the saddle point and can then transfer to the other
superconducting state.
When the magnetic
field approaches the expulsion or penetration field, the
attraction region of a saddle point state decreases and random searching
becomes inefficient. Therefore, to trace the saddle point evolution in 
the vicinity of the penetration (expulsion) field we start from
the saddle point state and increase (decrease) the magnetic field up
to the penetration (expulsion) field,
when the lowest eigenvalue goes to zero (see Fig.~\ref{fig2}, dashed curve).

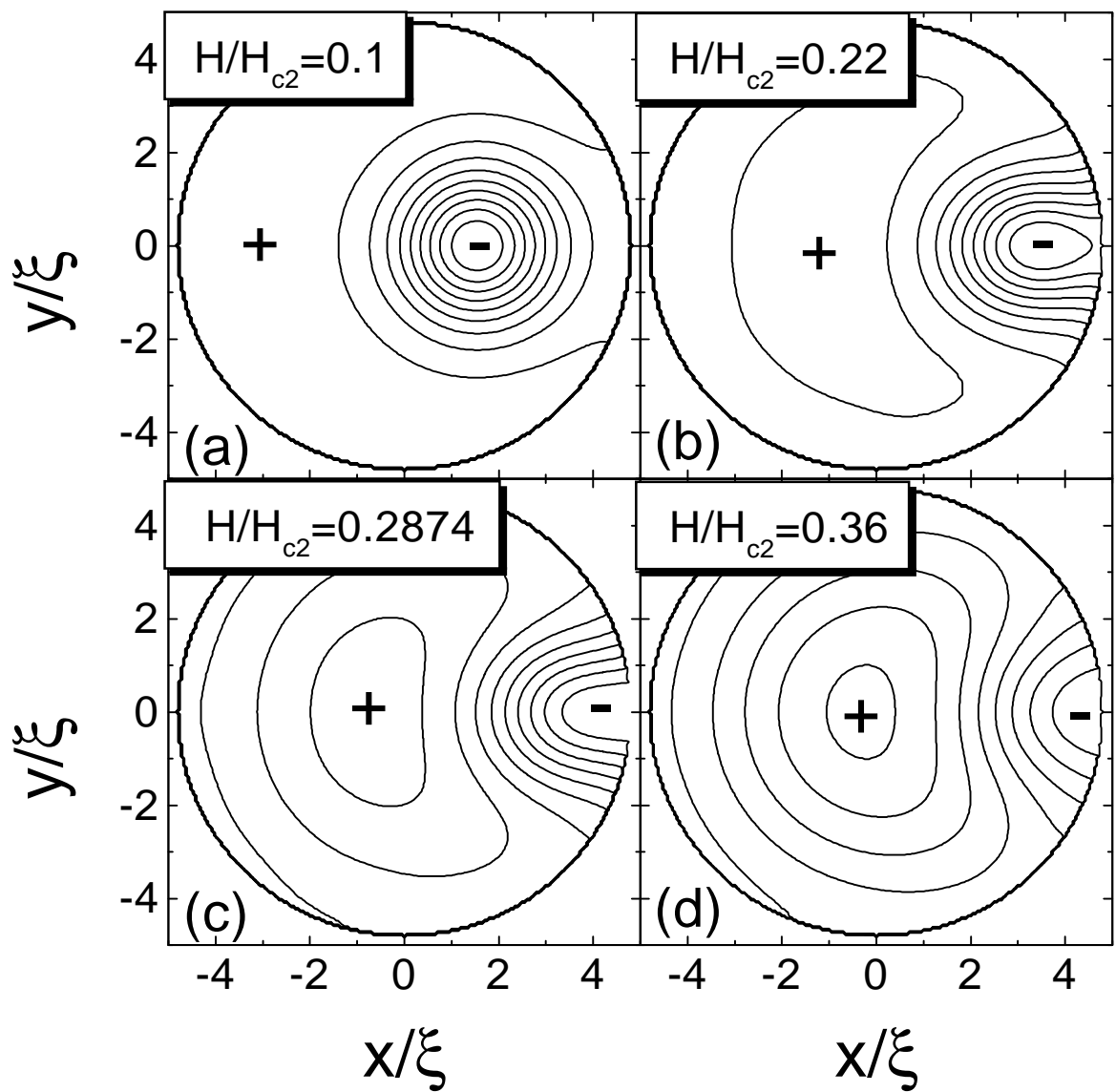
\begin{figure}
\caption{
Density of superconducting electrons $|\Psi|^2$ in the saddle
point states corresponding to the transition $0\leftrightarrow 1$
for the disk radius $R=4.8\xi$ and different magnetic fields.
The signs (+), (-) denote the maxima and minima of $|\Psi|^2$, respectively.}
\label{fig3}
\end{figure}

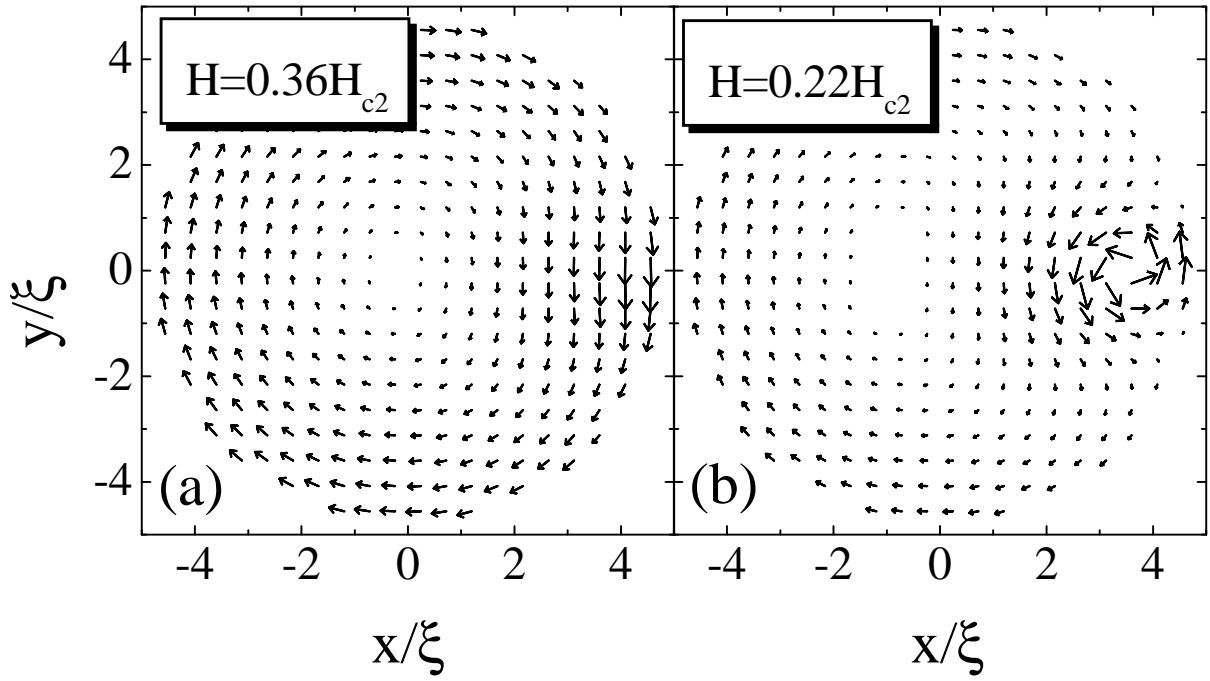
\begin{figure}
\caption{
The velocity of superconducting electrons 
in the saddle
point states corresponding to the transition $0\leftrightarrow 1$
for the disk radius $R=4.8\xi$ and for two different magnetic fields.
}
\label{fig4}
\end{figure}

The spatial distributions of the superconducting electron density $|\Psi|^2$
and velocity $\vec V=\hbar\vec \nabla S-2e\vec A/c$ ($\Psi=Fexp(iS)$)
in the saddle point state corresponding to the transition from the Meissner
state to the vortex state are depicted in Figs.~\ref{fig3} and \ref{fig4}
for different magnetic
fields and  disk radius $R=4.8\xi$. These figures demonstrate
two different stages in the saddle point evolution.
Below the penetration field, the
saddle point state corresponds to a region of suppressed
superconductivity (Fig.~\ref{fig3}(d)) with a minimum
of $|\Psi|^2$, which is located at
the disk boundary. While the minimum value of the order parameter
remains different from zero, the vorticity
$L=\int d\vec l\partial S/\partial \vec l/2\pi$, where integration is
performed along the disk boundary, equals zero and the supervelocity
distributions is similar to that of the Meissner state (Fig.~\ref{fig4}(a)).
When the order parameter reaches its zero value at  the {\it
nucleation field} $H_n$ (Fig.~\ref{fig2}, solid curve),
the vorticity transits suddenly to $L=1$. For lower magnetic fields $H<H_n$,
the saddle point state presents a
vortex-like state with closed velocity circulation (Fig.~\ref{fig4}(b)).
Note, that this transition 
is not followed by any discontinuity in the free energy or 
the curvature of the potential curve $\eta$. 
With further
decreasing the magnetic field, the saddle point corresponds to a vortex 
closer to the disk center (Fig.~\ref{fig3}(d),(c)).
This physical picture of flux
expulsion and penetration remains valid for other transitions
$L\leftrightarrow L+1$ with different $L$, independently of the disk
radius and the type (giant vortex or multivortex\cite{schweigert2})
of superconducting state.

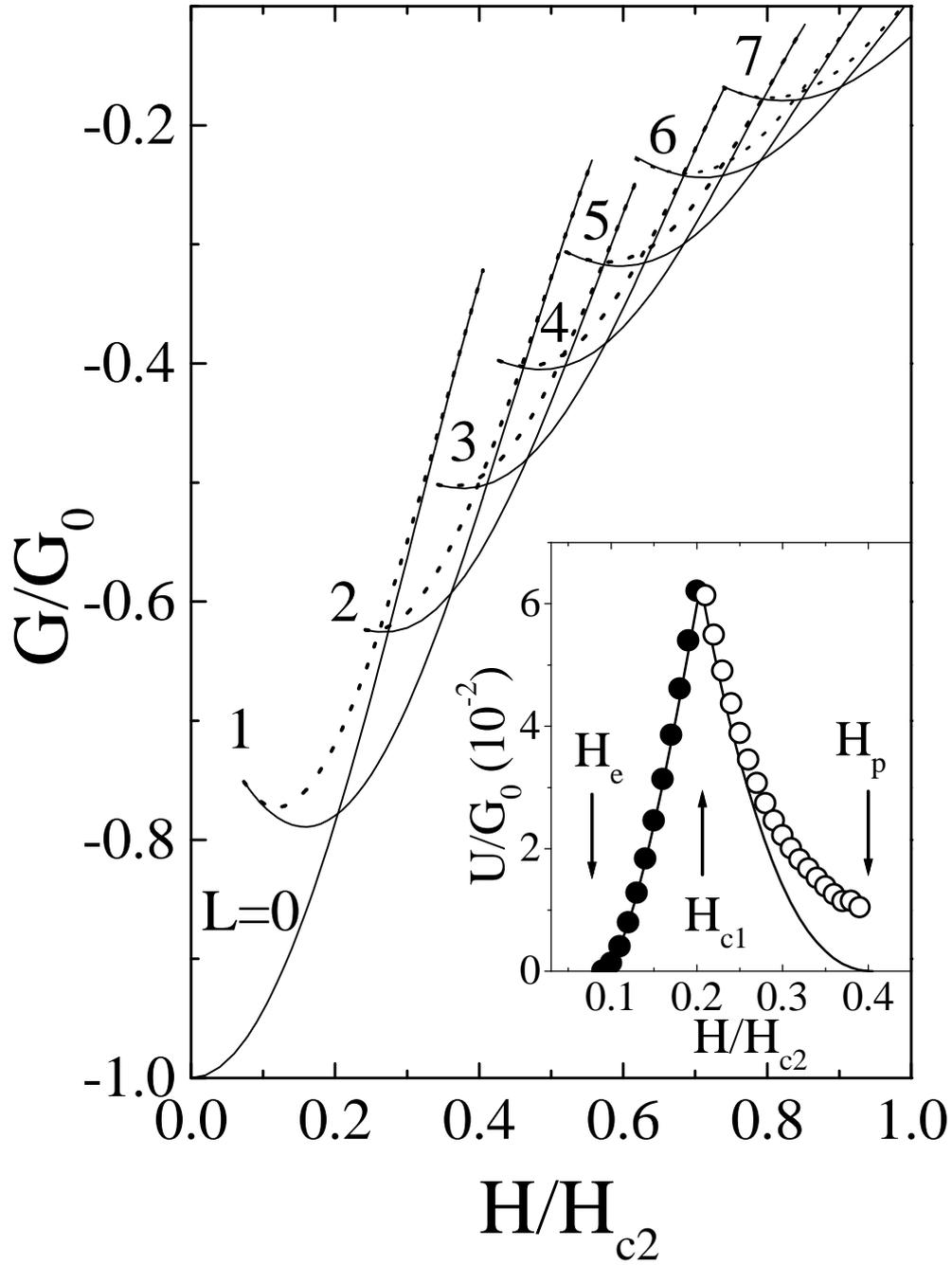
\begin{figure}
\caption{
The free energy of the stable (solid) and saddle point (dotted
curves) states as function of the magnetic field, measured in
$H_{c2}=\Phi_0/2\pi \xi^2$,
for different angular momenta $L$. The inset shows the expulsion
(solid circles)
and penetration (open circles) barriers
found with our modified Bean-Livinston model
(symbols) which are compared to our GL approach (curves)
for the transition $0\leftrightarrow 1$.}
\label{fig5}
\end{figure}

The free energy,
measured in the condensation energy $G_0=\alpha^2\pi R^2d/2\beta$,
is shown in Fig.~\ref{fig5}
for the saddle point (dotted curves) and stable (solid curves) states 
for the disk radius $R=4.8~\xi$.
The difference between the free energy
of the saddle point state and the nearby metastable state corresponds
to
the transition barrier shown in the inset of Fig.~\ref{fig5} for
transition $0\leftrightarrow 1$. 
 As seen from Fig.~\ref{fig5}, the penetration barrier grows more slowly
deep inside  the metastable region than the expulsion barrier.
Therefore, we expect a larger fluctuation of the
penetration field at a finite sample temperature,
which agrees with recent
experimental observations by Geim \cite{geim}.

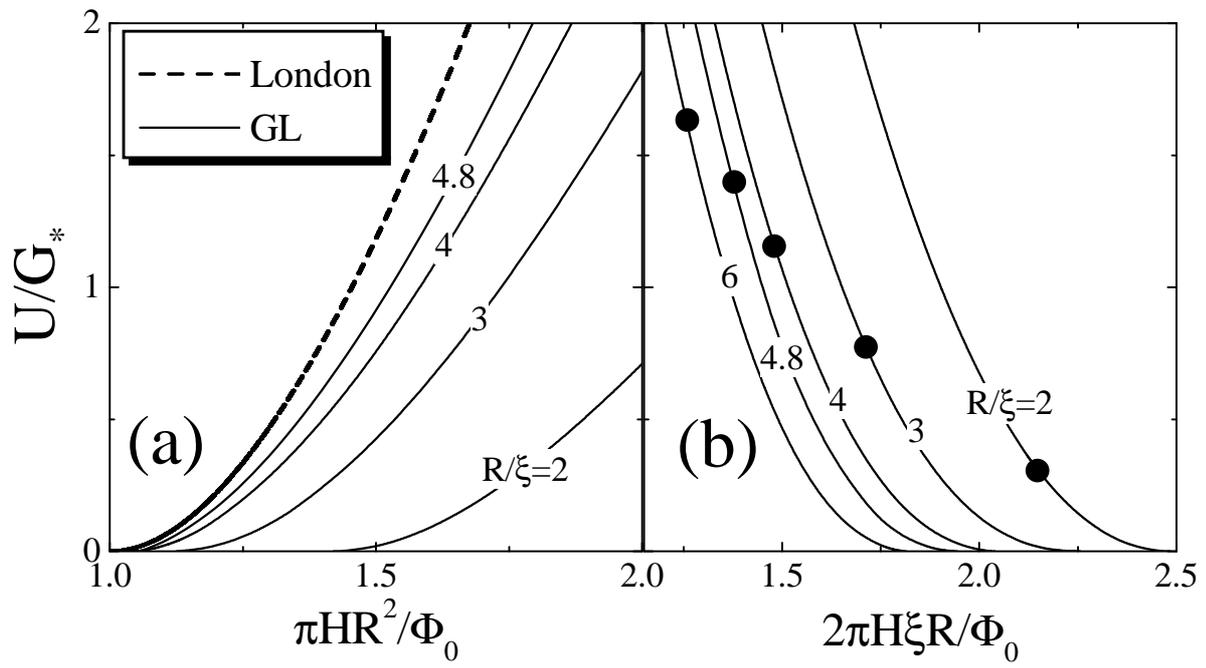
\begin{figure}
\caption{
The expulsion (a) and penetration (b) barrier as function of the
magnetic field for different disk radii. The dashed curve
corresponds to the London limit for the expulsion barrier. The solid
circles correspond to the penetration barrier at the nucleation field.
}
\label{fig6}
\end{figure}

Below the nucleation field, when the saddle point state is similar to
a vortex state, the
penetration and expulsion barriers (see Fig.~6) can be estimated
from the London theory, which leads to the
following expression for the vortex free energy \cite{buzdin}
\begin{equation}
\label{eq1}
\frac{G_1}{4\pi G_{\star}}=
ln(\frac{R^2-\rho^2}{r_c R})-\Phi (1-\frac{\rho^2}{R^2})
+\frac{1}{4}(\Phi^2-\frac{R^2}{\xi^2}),
\end{equation}
where $G_{\star}=\alpha^2\xi^2d/2\beta=G_0\xi^2/\pi R^2$,
$\rho$ is the radial vortex position, $r_c\sim \xi$ is the vortex core
radius, $\Phi=\pi HR^2/\Phi_0$ is the
unitless magnetic flux, and $\Phi_0=hc/2e$ is the flux quantum. Note
that: i) the expulsion field $H_e=\Phi_0/\pi R^2$, ii) the vortex position
$\rho_s=R\sqrt{1-1/\Phi}$ in the saddle point,
and iii) the BL  expulsion barrier $U=G_1(\rho_s)-G_1(0)=
4\pi G_{\star}(\Phi-1-ln\Phi)$ do not depend on the vortex core energy, which
is represented  by the first term in Eq.~(\ref{eq1}). 
As seen from Fig.~\ref{fig6}(a),
the London theory prediction for the expulsion barriers
are confirmed by the GL theory in the limit of large disks $R\gg \xi$.
We extended the BL model to arbitrary $R/\xi$ by taking into
account the spatial
nonuniformity of the modulus of the order parameter, which obeys
the first GL equation
\begin{equation}
\label{final}
-\frac{\hbar ^2}{2m^*}\left
(\triangle F-(\vec\nabla S-\frac{2e}{\hbar c}\vec A)^2F\right)=-\alpha F+\beta F^3,
\end{equation}
with the boundary condition $(\partial F/\partial \rho)_{\rho=R}=0$.
Following the BL model we
assume that the phase distribution is created by a vortex and its mirror image,
which are located at the distances $\rho_v, R^2/\rho_v$ from the disk center
\cite{fetter,buzdin}, respectively. Solving  Eq.~(\ref{final})
numerically
for different vortex positions $\rho_v$ we find the expulsion (solid circles)
and penetration (open circles) barriers
shown in the inset of Fig.~\ref{fig5}. Below the nucleation
field there is an
excellent quantative agreement between our GL theory and this improved
BL model. Nevertheless, this model breaks down 
in the range $H_n<H<H_p$.  Note, that the barrier height at
$H=H_n$ increases with disk radius and the role of {\it the
nucleation barrier} may become even more important in macroscopic systems,
where possible 3D (for $d>\xi$) and demagnetization (for
$Rd>\lambda^2$) effects must also be taken into account.
In the unitless variables $2\pi H\xi R/\Phi_0$ ($\Phi/\Phi_0$), the
penetration (expulsion) barriers measured
in $G_{\star} $ are proportional to the disk
thickness and increase slightly with the disk radius.

In conclusion,  we have  demonstrated
that  in a wide range of  magnetic fields $H_n<H<H_p$ the saddle
point state presents {\it a vortex nucleus}, which is a
 region of suppressed superconductivity surrounded by a
background of
Meissner state, which transits to a vortex  state at $H<H_n$.
We have found the
penetration field and the corresponding {\it nucleation barriers} for
thin disks. For lower magnetic fields $H_e<H<H_n$, the saddle point state can 
be reasonably described by the conventional London theory. We also
extended the BL model to finite disk radius.

We gratefully acknowledge discussions with A.K.~Geim, R.~Blossey,
A. MacDonald and V. Moschchalkov.
This work is supported by the Flemish Science Foundation (FWO-Vl)
through project 5.0277.97 and the
``Interuniversity Poles of Attraction Program - Belgian State, Prime Minister's
Office - Federal Office for Scientific, Technical and Cultural Affairs''.
One of us (FMP) is a research director with the FWO-Vl.

\end{document}